\documentclass[%
 reprint,
 amsmath,amssymb,
 aps,
]{revtex4-2}

\usepackage{physics}
\usepackage{amsmath}
\usepackage{booktabs}
\usepackage{siunitx}
\usepackage{easy-todo}
\usepackage{xfrac}
\usepackage{enumitem}

\usepackage{lineno}
\begin{document}
\title{Experimental demonstration of a simple free-space loss model for diffuse reflections in guided-mode LiDAR}

\author{Quentin Baudenon}
\affiliation{Baraja Pty. Ltd., North Ryde, NSW 2113, Australia}

\author{Ben Hopkins}
\email{ben.hopkins@baraja.com}
\affiliation{Baraja Inc., Palo Alto, CA 94306, USA}

\author{Cibby Pulikkaseril}
\affiliation{Baraja Pty. Ltd., North Ryde, NSW 2113, Australia}

\begin{abstract}
The growing use of photonic integrated circuits in LiDAR sensors has made the speckle and phase distribution of the reflected light relevant for coupling into the guided-modes of these systems. 
In this Article, we adopt the use of a simplified diffuse surface model to show the average coupling efficiency of reflected speckle into guided modes can be predicted from the optical intensity at the target. 
For the first time, we provide experimental results to show that this relationship is valid for predicting free-space loss as a function of range.
This Article is intended to aid optics design for integrated photonic LiDAR systems, providing a simple and accurate model to predict the free-space loss from a representative target at all distances.  
\end{abstract}


\maketitle

\section{Introduction}

\begin{figure}[t]
\centering
\fbox{\includegraphics[width=0.98\linewidth]{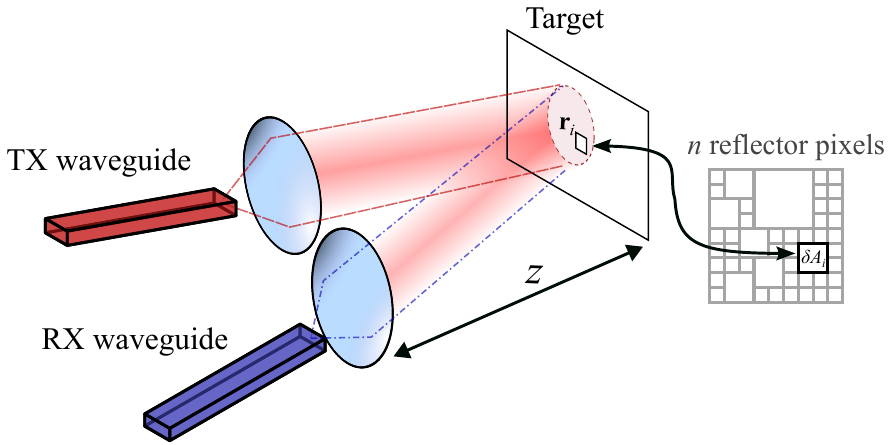}}
\caption{
An illustration of a biaxial \mbox{LiDAR} configuration, where transmit (TX) and receive (RX) waveguides are collimated by separate optical systems. 
The overlap of TX and RX beams occurs on a target at distance $z$, where we model the target surface as being made of $n$ reflector pixels that each have reflectivity $\rho$ and generate phase delays $\psi$.  
}
\label{fig:Schematic}
\end{figure}

\mbox{LiDAR} (Light Detection And Ranging) is an active sensor consisting of an optical transmitter that sends optical energy into the environment and an optical receiver that collects a fraction of the reflected energy, using the time delay between transmit and receive to calculate range.
Our interest lies in the application of \mbox{LiDAR} to autonomous navigation systems, where a large solid angle field-of-view is required and typically achieved through means of scanning or flash \mbox{LiDAR}.   
Scanning LiDAR uses a single transmitter and receiver pair that mechanically scans the desired field-of-view, while flash \mbox{LiDAR} uses a static grid of transmitters with another grid of receivers to illuminate and then detect over a field-of-view simultaneously.
Within these implementations there is also variation whether the optical link is achieved by: (i) pulsed time of flight, transmitting a single pulse of light, (ii) random modulated continuous wave, transmitting an amplitude or phase modulated codeword, or (iii) frequency modulated continuous wave, transmitting a chirped frequency pulse.
Details on the types of \mbox{LiDAR} implementations for autonomous navigation can be found in reviews on the subject, such as by Schleuning et al.~\cite{schleuning2020LiDAR}
In this Article, we are only concerned with the design of the optical collimation systems and the efficiency with which transmitted power returns to the associated receiver.  
This efficiency is common to all the \mbox{LiDAR} implementations listed, including coherent and non-coherent receiver architectures.  
Our specific interest is in scanning \mbox{LiDAR} using a single transmitter and receiver pair, and what we present is biased towards that implementation.  
With this in mind, a conceptual illustration of the optical collimation system used in a scanning \mbox{LiDAR} is shown in Fig.~\ref{fig:Schematic}, where the transmit (TX) and receive (RX) optical paths use separate collimating optics, otherwise known as a biaxial configuration.

For \mbox{LiDAR} applied to autonomous navigation, the system's probability of object detection over range is a key limiting requirement. 
Being able to predict this range performance of a LiDAR system is essential when developing its optical subsystems because it allows the designer to quantify optical performance relative to size and complexity of the optics. 
The incumbent method to estimate the received optical power is the {\it \mbox{LiDAR} Equation}~\cite{fersch2017challenges}, which is typically constructed for systems with large area detectors that convert aggregate optical power at the receiver into an average photocurrent.  
However, the \mbox{LiDAR} Equation does not account for key effects such as random field variation from laser speckle~\cite{baumann2014speckle} on the complex optical field overlap that exists when coupling into guided-modes.  
More details on the \mbox{LiDAR} Equation are provided in Appendix~\ref{apnd:LiDARequation}.
However, guided-mode LiDAR is a much closer analogue to coherent LiDAR or heterodyne receivers, where the usable signal is limited to that which overlaps with local oscillator fields~\cite{Siegman66}. 
For modeling coherent LiDAR specifically, the average received optical signal from a uniform reflectivity diffuse surface can be predicted from the overlap of intensities of the transmitter fields and back-propagated local oscillator fields at the target~\cite{Frehlich1991}. 
In fact, this method is referred to as the back-propagated local oscillator (BPLO) approach, and has been widely used for predicting the performance of coherent atmospheric LiDAR sensors~\cite{Zhao1990}.
However, the BPLO method was developed to predict the received signal from distant targets; the intensity overlap result was derived under the Fresnel Approximation.~\cite{Zhao1990, Frehlich1991} 
The existing literature also does not experimentally investigate the accuracy of the BPLO method over range.
The utility of any model for the application of guided-mode LiDAR to autonomous navigation systems requires strong evidence that it matches reality, and that it can accurately predict detection performance for both close and distant targets. 
To the best of the authors' knowledge, it has not been shown that the intensity overlap result from coherent LiDAR can be used generally to cover both close and distant ranges.

In this Article, we employ a basic model for the diffuse surface instead of making assumptions about the optical propagation or the range, which allows us to dramatically simplify the calculation of the average free-space loss for a \mbox{LiDAR} sensor with guided-mode receivers.
Using this surface model, we are able to derive that the average free-space loss of a guided-mode LiDAR is proportional to the intensity overlap of transmitter and receiver fields at the target at all ranges.   
We then present experimental measurements of the free-space loss from biaxial and coaxial LiDAR transceivers over range, and see very good agreement with the predictions of our model.   
When compared to either the BPLO method or the LiDAR Equation, the key feature of the method we present is that it is built without assumptions about the propagation and it should thereby remain valid at all ranges.   


\section{Model}

The basis of our model is to treat the scattering surface as a distribution of independent reflectors that each apply a random phase shift and polarization change to the reflected light, which we will refer to as reflector pixels and are depicted in Fig.~\ref{fig:Schematic}. 
This surface model was chosen to resemble the arrays of coherent radiators that generate the monochromatic speckle patterns studied in the work of Ochoa and Goodman~\cite{ochoa1983statistical}.   
The array of reflector pixels is also a normal-incidence approximation of both Lambertian reference materials, such as Labsphere's Spectralon\textsuperscript\textregistered, and to existing models of diffuse scattering surfaces as distributions of V-shaped cavities~\cite{oren1994generalization, torrance1967theory}.
Spectralon\textsuperscript\textregistered~randomizes and depolarizes the reflected phase front from multiple internal reflections~\cite{bhandari2011bidirectional}, while normal incidence rays reflected from random V-shaped cavities generate a flat phase delay across each such cavity, which is dependent on the depth and width of the cavity.
To align with these two examples, we only consider situations where both the angle of illumination and the angle of observed reflection are normal to the target scattering surface.

The coupling efficiency between transmit and receive ports of a \mbox{LiDAR} system can be computed as the overlap integral of receiver’s port modes with the reflected beam profile, evaluated on any surface that contains the intersection of the field profiles. However, computing the overlap integral at the RX waveguide interface requires simulation of the reflected speckle pattern passing through the \mbox{LiDAR} system optics.  Instead, as done by Frehlich and Kavaya~\cite{Frehlich1991}, we propagate both the TX and RX scalar electric field profiles, $U_{\scriptscriptstyle \text{TX}}$ and $U_{\scriptscriptstyle \text{RX}}$, out to the target and perform the overlap integral at the target. 
The coupling efficiency, $\eta$, can then be written as an overlap of the Rx mode onto the reflection of the TX mode: 
\begin{equation}
\label{eq:eta_definition}
    \eta = \abs{ \int{ U^*_{\scriptscriptstyle \text{RX}} \cdot \left( \sqrt{\rho} e^{j\psi} \cdot U_{\scriptscriptstyle \text{TX}} \right) dA} }^2 
\end{equation}
where $\rho$ is the target reflectivity, $\psi$ is the phase introduced by the reflector pixels, and the scalar electric fields, $U_{\scriptscriptstyle \text{TX}}$ and $U_{\scriptscriptstyle \text{RX}}$, are normalized such that $\int{\abs{U}^2 dA} = 1$ at their respective waveguide ports. 
Note that $U_{\scriptscriptstyle \text{TX}}$ and $U_{\scriptscriptstyle \text{RX}}$ evaluated at the target will include transmission losses such that $\int_{target}{\abs{U}^2 dA} = \zeta$, where $\zeta$ is the one-way transmission efficiency.
It is also worth recognizing that by reducing a vector field overlap to a scalar field overlap, we have assumed that the polarization of both the reflected transmission and the back-propagated receiver fields lie in the same plane, and that the receiver has degenerate polarizations to account for any in-plane polarization rotation or birefringence from reflection.  
This reduction will notably not be true for grazing angles of incidence on the target.  
Continuing on the premise that we are considering only normal incidence, we write out \eqref{eq:eta_definition} as a product of two integrals
\begin{align}
    \eta = \rho & \cdot
\left( \int{U^*_{\scriptscriptstyle \text{RX}}(\textbf{r}) \cdot U_{\scriptscriptstyle \text{TX}}(\textbf{r}) \cdot e^{j\psi} dA} \right)
   \nonumber \\ & \cdot
\left( \int{U_{\scriptscriptstyle \text{RX}}(\textbf{r}') \cdot U^*_{\scriptscriptstyle \text{TX}}(\textbf{r}') \cdot e^{-j\psi '} dA'} \right)
\end{align}
After expanding into a single double integral, and then treating the phase of each reflector pixel as an independent random variable, the expected coupling efficiency over these random variables is given by
\begin{align}\label{eqn:expected_eta}
E[\eta] = \rho \cdot \iint & U^*_{\scriptscriptstyle \text{RX}}(\textbf{r}) \cdot U_{\scriptscriptstyle \text{TX}}(\textbf{r})  \cdot U_{\scriptscriptstyle \text{RX}}(\textbf{r}') \cdot U^*_{\scriptscriptstyle \text{TX}}(\textbf{r}') 
\nonumber \\
           &
\cdot E[e^{j(\psi-\psi ')}] dA \,  dA'
\end{align}
where $E[\cdot]$ is the expected value of a stochastic process.
As the phase of each reflector pixel, $\psi$, is assumed to be a random uniform distribution over $[0,2\pi)$ in the reflector pixel surface model, we can therefore make the simplification
\begin{equation}
   E[e^{j(\psi - \psi ')}]  = \begin{cases}
  1  & \text{if } \psi = \psi ' \\
  0 & \text{otherwise}
\end{cases}
\end{equation}
In other words, the integrand in \eqref{eqn:expected_eta} is non-zero if and only if positions $\textbf{r}$ and $\textbf{r}'$ lie on the same reflector pixel. Contributions from all other pairs of reflector pixels have an expectation value of 0. 
To simplify the calculation further, we now assume that the reflector pixels are small enough that the fields $U_{\scriptscriptstyle \text{TX}}$ and $U_{\scriptscriptstyle \text{RX}}$ are constant over the area of the pixel. If there are $n$ reflector pixels at positions $\textbf{r}_i$ with areas $\delta A_i$, \eqref{eqn:expected_eta} becomes
\begin{equation}
\label{eqn:FSL_nearlyfinal}
    E[\eta] \approx \rho \cdot \sum^n_{i=1} \abs{U_{\scriptscriptstyle \text{RX}} (\mathbf{r}_i )}^2 \cdot \abs{U_{\scriptscriptstyle \text{TX}}(\mathbf{r}_i)}^2 \, (\delta A_i)^2
\end{equation}
Assuming all reflector pixel areas are equal, $\delta A_i \equiv \delta A$, a single $\delta A$ can be brought outside of the sum over $i$ as a common factor, which makes the remaining sum an area integral of
$\abs{U_{\scriptscriptstyle \text{RX}}}^2 \cdot \abs{U_{\scriptscriptstyle \text{TX}}}^2$.   
In this manner, $E[\eta]$ simplifies to the product of a design-dependent characteristic, the overlap integral of the intensities $I_{\scriptscriptstyle \text{TX}}$ and $I_{\scriptscriptstyle \text{RX}}$ (where $I = |U|^2$), and a target-dependent characteristic, $\rho \cdot \delta A$
\begin{equation}\label{eqn:FSL_final}
    E[\eta] \approx \rho \cdot \delta A \cdot \int I_{\scriptscriptstyle \text{TX}} \cdot I_{\scriptscriptstyle \text{RX}} \; dA
\end{equation}
Here \eqref{eqn:FSL_final} demonstrates a surprising conclusion: though we are coupling a speckled beam into our receiver, which is inherently a phenomenon of field interference, the stochastic expected coupling doesn't depend on the field phase profile as it is calculated only as the overlap of the intensities of the TX and RX beams.
The practical conclusion of \eqref{eqn:FSL_final} for a given target surface, is that any expected free-space loss variation will be entirely predictable by variation in the TX and RX intensity profiles at the target.  
For instance, this means that the dependence of expected loss on distance to the target, $z$, is contained in how the intensity profiles evolve with $z$.  
It also means that the impact of aberrations can be predicted provided accurate intensity profiles of the aberrated beams at the target.
Along comparable lines, if the given LiDAR system has a laser source with some linewidth, we would calculate $E[\eta]$ as a function of frequency then use it as a linear operator on the distribution of transmission power over frequency, which allows us to determine the total received power over the linewidth.  
This means the variation of the intensity profiles over the relevant frequency range of the laser, will dictate for instance whether we can approximate the laser linewidth as being a single frequency.
Similar analysis could also be applied to model the effect of doppler shifts caused by vibration.  
In the next section, we will demonstrate that the expression in \eqref{eqn:FSL_final} can accurately predict the free-space loss of a diffuse scattering target at normal incidence.  

\section{Measurements}
\label{sec:measurments}

In this section, we provide experimental validation of the free-space loss model in \eqref{eqn:FSL_final} for a reference target at normal incidence.
This validation is presented in two parts: the experiment in~\ref{sec:model_calibration} estimates $\delta A$ for a reference target by fitting measured data to \eqref{eqn:FSL_final}, while the experiment in~\ref{sec:validation} validates whether this $\delta A$ and \eqref{eqn:FSL_final} can be used with numeric simulations to accurately predict the free-space loss for the same reference target and different transceivers.   

\subsection{Calibration Experiment}\label{sec:model_calibration}
\begin{figure}[ht]
\centering
\fbox{\includegraphics[width=0.98\linewidth]{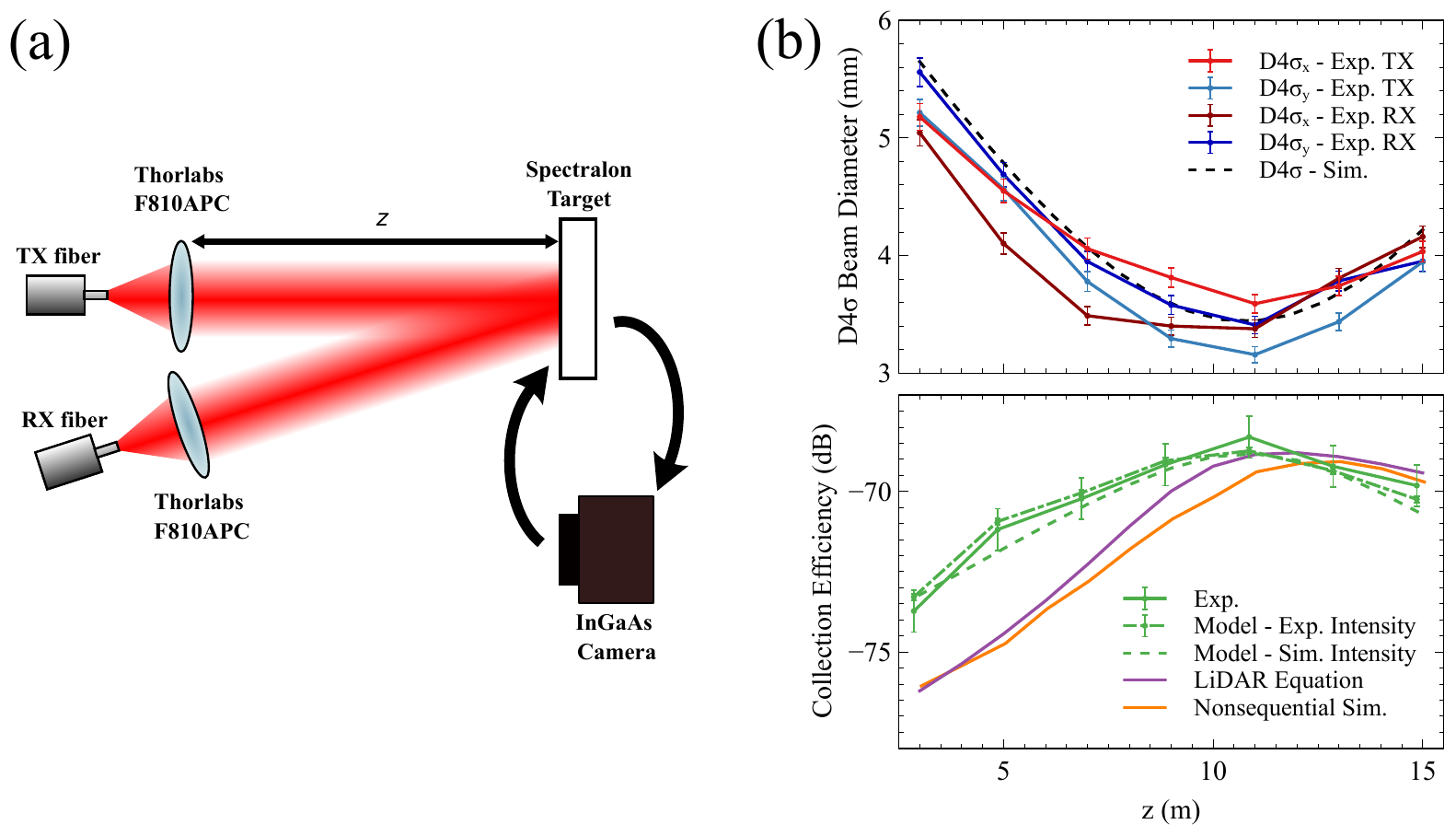}}
\caption{(a) Experimental setup of a biaxial LiDAR transceiver used to measure beam intensity profiles and free-space loss from a Spectralon\textsuperscript\textregistered~target, for distances from \SI{3}-\SI{15}{\meter}.  
The TX and RX beam diameters in (b, top) were measured on an InGaAs camera (MKS Ophir SP1201) with a \SI{1550}{\nano\meter} laser source connected to the TX and then RX fiber.  
The free-space loss from the Spectralon\textsuperscript\textregistered~target, labeled as \mbox{\it Exp.} in (b, lower), was measured with a \SI{1550}{\nano\meter} laser source connected to the TX fiber and an optical power meter connected to the RX fiber.   
The curve labeled \mbox{\it Model - Exp. Intensity} was created by inserting the measured beam intensity profiles into \eqref{eqn:FSL_final} with $\delta A$=\SI{1.20}{\micro\meter^2}.  
Multiple approaches to predict free-space loss from simulation using Zemax OpticStudio are also overlaid in (b, lower) and discussed in Section~\ref{sec:discussion}.   
\mbox{\it Model - Sim. Intensity} is from \eqref{eqn:FSL_final} using simulated intensity profiles with $\delta A$=\SI{1.20}{\micro\meter^2} (details in Appendix~\ref{apnd:SimFSL}), \mbox{\it Nonsequential Sim.} is from nonsequential ray tracing with a Lambertian target, and \mbox{\it LiDAR Equation} is from \eqref{eq:LiDAREqnCE} using the \mbox{LiDAR} Equation for a Lambertian target (details in Appendix~\ref{apnd:LiDARequation}).
}
\label{fig:model_validation}
\end{figure}

To estimate $\delta A$  in \eqref{eqn:FSL_final} for a given target surface, we conducted a simple experiment with a biaxial transceiver that is composed of two identical doublet fiber collimators (Thorlabs F810APC-1550), each connected to an FC/PC single mode fiber. 
As illustrated in Fig.~\ref{fig:model_validation}(a), this biaxial transceiver was aimed at a Lambertian reference material (Spectralon\textsuperscript\textregistered, 99\%). 
The TX collimator's fiber was connected to a continuous wave laser (Coherent Solutions \mbox{IQTLS-1-C-S-FA}) set to a wavelength of \SI{1550}{\nano\meter} and a linewidth of \SI{100}{\kilo\hertz}.  
The RX collimator's fiber was connected to an optical power meter to measure the reflected power.
Prior to this, the one-way transmission efficiency was also recorded for both TX and RX collimators, $\zeta_{\scriptscriptstyle \text{TX}}$ and $\zeta_{\scriptscriptstyle \text{RX}}$, using a free-space optical power meter placed immediately after each Thorlabs F810APC doublet collimator.  

The measurements of free-space loss were then performed by placing the Spectralon\textsuperscript\textregistered~reference target at multiple distances between \SI{3}{} and \SI{15}{\meter} from the transceiver, while keeping the TX beam normal to the target.  
At each distance, the RX collimator was re-aligned in pitch and yaw to maximize returned power. 
This realignment was to ensure the TX and RX beams were maximally overlapping for every distance, and avoided creating a blind spot between the two collimators. 
The existence of speckle in the reflection does cause large variations in the power measured through the RX collimator, depending on the illuminated area of our target. 
To average these variations, the power measurements were averaged over \SI{30}{\second} while the target was rotating at \SI{0.067}{\hertz} about a rotation axis that is normal to the surface. 
This rotation axis was transversely offset relative to the illuminating beam so different parts of the surface are illuminated during the \SI{30}{\second} averaging window, while maintaining normal incidence of TX beam.
Using the laser's transmit power, $P_{\scriptscriptstyle \text{TX}}$, and the measured average received power in the RX fiber, $\overline{P_{\scriptscriptstyle \text{RX}}}$, the free-space loss' contribution to coupling efficiency is then defined as:
\begin{equation}
\label{eq:FSL_exp}
E[\eta_{\scriptscriptstyle \text{FSL}}] \equiv  \frac{E[\eta]}{\zeta_{\scriptscriptstyle \text{TX}} \cdot \zeta_{\scriptscriptstyle \text{RX}} \cdot \rho} = \frac{\overline{P_{\scriptscriptstyle \text{RX}}}/P_{\scriptscriptstyle \text{TX}}}{ \zeta_{\scriptscriptstyle \text{TX}} \cdot \zeta_{\scriptscriptstyle \text{RX}} \cdot \rho}
\end{equation}
This free-space loss measurement is labeled as {\it Exp.}~in Fig.~\ref{fig:model_validation}(b,lower).
After the free-space loss measurement at each target distance, the Spectralon\textsuperscript\textregistered~target was removed and replaced by an \mbox{InGaAs} camera (MKS Ophir SP1201) to measure the optical intensity profile from the TX collimator at that distance.
Without moving the camera, we then connected the laser to the RX collimator and similarly measured the optical intensity profile from the RX collimator.  
These measurements provided us with the two-dimensional optical intensity profiles of both the TX and RX beams over range, which we could use in the prediction expression for the expected free-space loss in~\eqref{eqn:FSL_final}.
Using the reflector pixel area $\delta A$ as a free variable, we get a good fit between the measured free-space loss and the expected free-space loss from these measured intensity profiles by using \mbox{$\delta A$=\SI{1.20}{\micro\meter^2}}.  
The curve of the expected free-space loss from measured intensity profiles and \mbox{$\delta A$=\SI{1.20}{\micro\meter^2}} is labeled \mbox{\it  Model - Exp. Intensity} in Fig.~\ref{fig:model_validation}(b,lower).   
The RMS error of the fit is 0.34~dB and is within the test setup’s error margin, which was estimated at $\pm0.65$~dB from a gauge repeatability and reproducibility analysis on the measurement system.
The coming experiment in~\ref{sec:validation} will show that this calibrated value of \mbox{$\delta A$=\SI{1.20}{\micro\meter^2}} allows us to accurately predict the free-space loss of different transceivers when using this same Spectralon\textsuperscript\textregistered~target.

\subsection{Prediction Experiment}
\label{sec:validation}

\begin{figure}[ht]
\centering
\fbox{\includegraphics[width=0.98\linewidth]{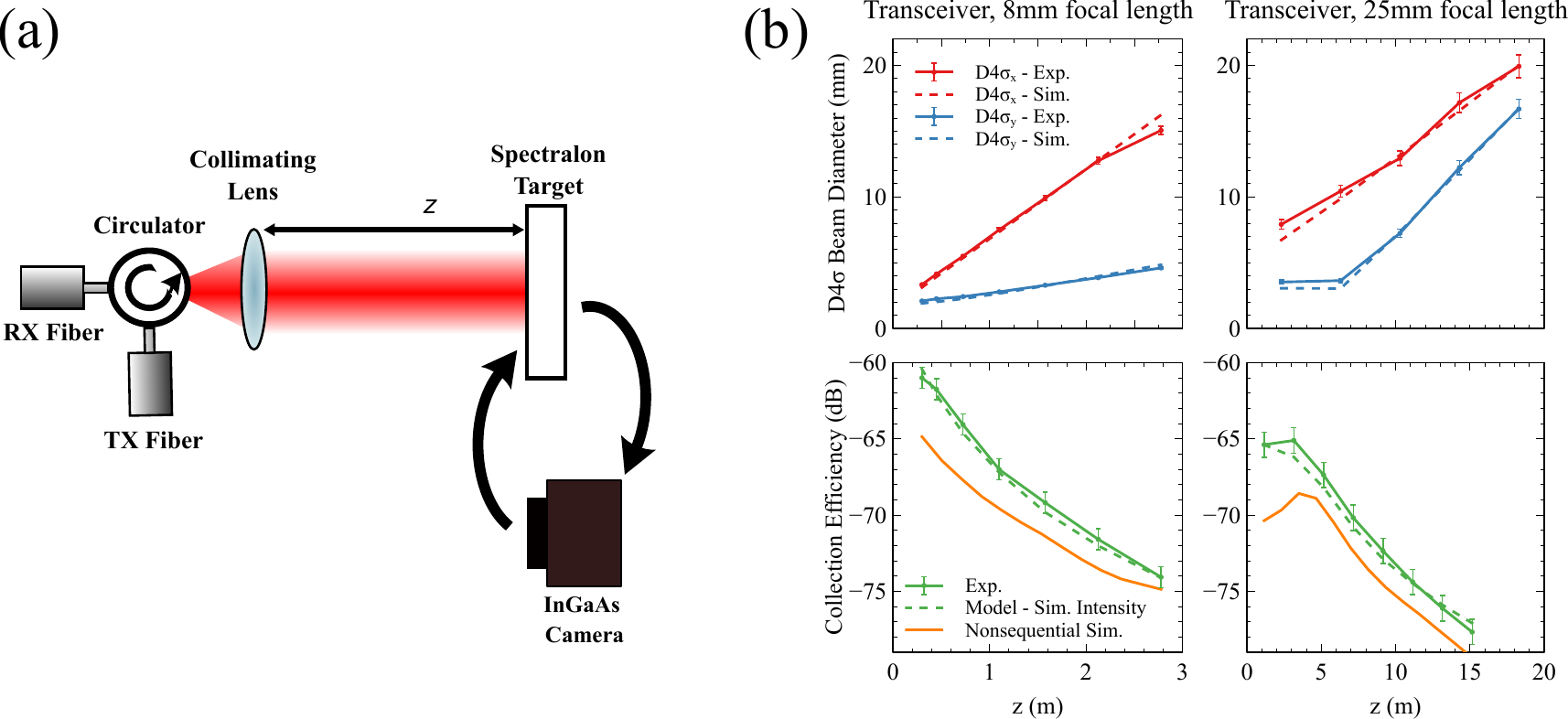}}
\caption{
(a) Experimental setup of a coaxial LiDAR transceiver used to measure beam intensity profiles and free-space loss from a Spectralon\textsuperscript\textregistered~target, for distances from \SI{0.3}-\SI{3}{\meter} with a Thorlabs 354240-C collimating lens (\SI{8}{\milli\meter} focal length) and \SI{3}-\SI{15}{\meter} with a Thorlabs AL1225-C collimating lens (\SI{25}{\milli\meter} focal length).  
The TX beam diameters in (b, top) were measured on an InGaAs camera (MKS Ophir SP1201) with a \SI{1550}{\nano\meter} laser source connected to the TX fiber.  
The free-space loss from the Spectralon\textsuperscript\textregistered~target, labeled as \mbox{\it Exp.} in (b, lower), was measured with a \SI{1550}{\nano\meter} laser source connected to the TX fiber and an optical power meter connected to the RX fiber.  
Two approaches to predict free-space loss from simulation using a Zemax model are also overlaid in (b,lower) for comparison.   
\mbox{\it Model - Sim. Intensity} uses simulated intensity profiles in \eqref{eqn:FSL_final} with $\delta A$=\SI{1.20}{\micro\meter^2} (details in Appendix~\ref{apnd:SimFSL}), and \mbox{\it Nonsequential Sim.} uses nonsequential ray tracing with a Lambertian target.
}
\label{fig:predictive}
\end{figure}

Using the result of the calibration experiment in Section~\ref{sec:model_calibration}, that \mbox{$\delta A$ = \SI{1.20}{\micro\meter^2}} represents the 99\% Spectralon\textsuperscript\textregistered~target, we performed a second experiment to verify whether we can predict the free-space loss from the same Spectralon\textsuperscript\textregistered~target for different \mbox{LiDAR} transceivers. 
We first prepared two coaxial transceivers using different collimation optics: one with an \SI{8}{\milli\meter} focal length lens (Thorlabs 354240-C) and the other with a \SI{25}{\milli\meter} focal length lens (Thorlabs AL1225-C).
Both transceivers are composed of a TX and an RX fiber connected to an optical circulator whose output port is collimated by a single collimating lens, depicted in Fig.~\ref{fig:predictive}(a). 
To reduce back reflections, collimating lenses needed to be tilted to an angle of $17^\circ$ off-normal.   
Analogous to the experiment in Section~\ref{sec:model_calibration}, the TX fiber was connected to the \SI{1550}{\nano\meter} continuous wave laser of power $P_{\scriptscriptstyle \text{TX}}$ and the RX fiber was connected to an optical power meter to measure the average returned power $\overline{P_{\scriptscriptstyle \text{RX}}}$ over \SI{30}{\second}. 
The optical power that is back reflected from the collimating lens into the RX fiber, $P_0$, was also measured once by placing an absorptive black target in front of the transceiver.  
The measurements of $\overline{P_{\scriptscriptstyle \text{RX}}}$ were then offset by $P_0$ to isolate the power returned after reflection from the Spectralon\textsuperscript\textregistered~target.  
Other than this offset, the free-space loss were calculated from measurements as per \eqref{eq:FSL_exp}.  

To now make predictions of the expected free-space loss, the TX and RX beam intensity profiles are simulated from a model of the transceiver in Zemax~OpticStudio using Physical Optics Propagation.
To build accurate models of both transceivers, the lens alignment in both transceivers were experimentally characterized by measuring the TX beam intensity profiles over range with an InGaAs camera (MKS Ophir SP1201).   
The associated beam diameters are shown as {\it D4$\sigma x$ - Exp.} and {\it D4$\sigma y$ - Exp.} in Fig.~\ref{fig:predictive}(b,top).  
To make the lens alignment in the simulations approximate that of the experimental transceivers, the lens defocus and decenter were adjusted until the beam diameter over distance in simulation resembled that observed in experiment, as shown with {\it D4$\sigma x$ - Sim.} and {\it D4$\sigma y$ - Sim.} in Fig.~\ref{fig:predictive}(b,top).  
The expected free-space loss was then computed from the simulated TX and RX intensity profiles using \eqref{eqn:FSL_final} and the calibrated value $\delta A$ = \SI{1.20}{\micro\meter^2} from the biaxial transceiver experiment.
The curves are labeled as {\it Model Sim. Intensity} Fig.~\ref{fig:predictive}(b,lower), and show good agreement to the free-space loss measured in the experiment. 
The RMS error between simulation and experiment is 0.2~dB for the \SI{8}{\milli\meter} focal length transceiver and 0.37~dB for the \SI{25}{\milli\meter} focal length transceiver, which is within the test setup error margin estimated at $\pm 0.69$~dB and $\pm 0.83$~dB during the measurement system gauge repeatability and reproducibility analysis. 
This result demonstrates that we are able to accurately predict the free-space loss from the reference surface over range using the prediction expression \eqref{eqn:FSL_final} with $\delta A$ = \SI{1.20}{\micro\meter^2}.

\section{Discussion}
\label{sec:discussion}
The measurements presented in Section~\ref{sec:measurments} also let us compare our method's predictions in \eqref{eqn:FSL_final} to the predictions of the \mbox{LiDAR}~Equation, using numerical simulations of the transceivers.
To build an accurate simulation model for the biaxial transceiver, we used an axially-symmetric Zemax~OpticStudio model of the Thorlabs F810APC doublet collimator and adjusted the defocus until the simulated beam diameter, labeled as {\it D4$\sigma$ - Sim.} in Fig.~\ref{fig:model_validation}(b,top), approximated that of the measured beam diameters for both TX and RX collimators.   
To confirm the accuracy of the simulation model for the biaxial transceiver, we also used the simulated intensity profiles over distance to predict the expected free-space loss with \eqref{eqn:FSL_final} and $\delta A$ = \SI{1.20}{\micro\meter^2}.
Details on how we have used simulated intensity profiles in \eqref{eqn:FSL_final} to predict free-space loss are provided in Appendix~\ref{apnd:SimFSL}.
We observed a close match of these simulation predictions to the measured free-space loss, see the curves labeled {\it Model - Sim. Intensity} and {\it Exp.} in Fig.~\ref{fig:model_validation}(b,lower).  
Using the same Zemax model, we then calculated the free-space loss predicted by the LiDAR~Equation, labeled as {\it LiDAR Equation} in Fig.~\ref{fig:model_validation}(b,lower).  
Details of the calculation are provided in Appendix~\ref{apnd:LiDARequation}, but the core assumptions are that the target is a Lambertian scatterer and we are approximating the fiber core as a photodetector with \SI{9}{\micro\meter} diameter.  
To validate the implementation of the LiDAR Equation, we then performed a nonsequential ray tracing simulation of the same transceiver model, using a \SI{9}{\micro\meter} diameter photodetector and a Lambertian scattering target.  
The \mbox{LiDAR} Equation can be seen to approximate the same curve as the nonsequential simulation, comparing the respective curves labeled as {\it Nonsequential Sim.} and {\it LiDAR Equation} in Fig.~\ref{fig:model_validation}(b,lower).   
However, we can see that the curves generated from our prediction expression using \eqref{eqn:FSL_final} provide a better estimate of the measured free-space loss. 
A similar conclusion can be made for the coaxial transceviers in Fig.~\ref{fig:predictive}(b,lower), where applying \eqref{eqn:FSL_final} gives more accurate predictions of free-space loss than nonsequential ray tracing.
There are no curves for the predictions from LiDAR Equation provided in Fig.~\ref{fig:predictive}, which is because the aberrations created by tilting the collimation lens to minimize back reflection makes the effective area calculation difficult and inaccurate.
That the method presented in this Article is providing more accurate predictions of free-space loss compared to the {LiDAR Equation} and nonsequential ray tracing is likely due to the differences caused by: modeling guided-mode receivers as large-area photodetectors, variations in propagation between wave optics and ray tracing, and the extent to which our Spectralon\textsuperscript\textregistered~target can accurately be treated as a Lambertian scattering surface.  
It is finally worth recognizing that \eqref{eqn:FSL_final} is similar to the prediction expression from the back propagated local oscillator (BPLO) approach for an infinite diffuse surface in Frehlich and Kavaya's work~\cite{Frehlich1991} (see Eq.~83).  
Though we have not used the Fresnel Approximation in our work and their work has modeled the diffuse surface as having a random reflection coefficient, both methods nonetheless predict that the free-space loss is proportional to the intensity overlap of transmitted and back-propagated receiver fields at the target.

We have presented only accurate predictions from \eqref{eqn:FSL_final} in this work, but we should recognize common scenarios where we expect to break assumptions about modeling the target surface as an array of reflector pixels and our other approximations.  
They are as follows:
\begin{itemize}
\item[i.]	
The phases induced by reflector pixels, the $\psi$ in \eqref{eq:eta_definition}, become correlated to each other or the transmitted beam.  
Physically, this could be from smoothness, order or homogeneity in the target's surface roughness, or if the surface feature size is small enough compared to the wavelength to behave as an effective index medium. 
\item[ii.]	
The reflector pixel areas, $\delta A_i$, are sufficiently large that we can no longer assume the field amplitude is constant over each reflector pixel in \eqref{eqn:FSL_nearlyfinal}, with the extreme example being a mirror target.   
\item[iii.]
The target surface or an oblique illumination angle requires a vector field overlap integral to accurately calculate coupling into the receiver mode, rather than the scalar field overlap we define in \eqref{eq:eta_definition}.  
\item[iv.]
The target surface reflectivity, $\rho$, and/or reflector pixel size, $\delta A_i$, cannot be validly approximated as a constant across the whole target surface.  More accurate descriptions of diffuse surfaces can generally make use of a distribution of pixel sizes and reflectivities~\cite{oren1994generalization}.
\end{itemize}
Many of these exceptions represent avenues where the approach presented in this Article could be extended, particularly in handling polarization and oblique illumination angles.  
However, the prediction expression in \eqref{eqn:FSL_final} was developed as a tool to aid the design of collimation optics in guided-mode LiDAR systems, as opposed to modelling free-space loss dependence over different target surfaces and illumination angles.  
The scope of the current work is therefore only to provide and demonstrate a simple means to make accurate estimates of free-space loss from a representative diffuse target.  

\section{Conclusions}
In this Article, we have derived and experimentally validated a simple numerical method
to predict the free-space loss of guided-mode \mbox{LiDAR} systems aimed at diffuse scattering targets
at normal incidence.
Our intent was to aid bulk optics design in guided-mode LiDAR systems by predicting the free-space loss for a representative target at {\it any} distance, where we assumed a basic model for the diffuse scattering surface using reflector pixels, rather than making assumptions about optical propagation or range.  
The resulting expected free-space loss expression is similar to that of the BPLO method for coherent atmospheric LiDAR~\cite{Frehlich1991}, in that it is proportional to the intensity overlap of transmitter and back-propagated receiver (or local-oscillator) fields at the target.
However, our derived expression is valid at any distance and includes the reflector pixel area as a target-dependent degree of freedom that can be calibrated to better represent the desired reference target.  
As this method only relies on simplifying assumptions being applied to the target surface, its accuracy is determined by the extent to which the diffuse surface model using reflector pixels is representative of the given physical surface.   
The experimental measurements have demonstrated our method can accurately predict the free-space loss of coaxial and biaxial transceivers illuminating a 99\%~reflectivity Spectralon\textsuperscript\textregistered~target over a range of~\SI{15}{\meter}.

\appendix

\section{Comparison to the \mbox{LiDAR} Equation}
\label{apnd:LiDARequation}

The \mbox{LiDAR} Equation is the incumbent method used to predict the received optical power in a LiDAR systems for autonomous navigation, with large area photodetectors~\cite{fersch2017challenges}. 
It combines the transmitted power and the fraction of reflected power that is geometrically projected onto the receiver's limiting aperture, with additional losses from the target, optics and environment. 
In the case of a Lambertian target without atmospheric absorption, the ratio of received to transmitted optical power in the LiDAR Equation, $\eta_{\scriptscriptstyle \text{LE}}$, can be simplified to
\begin{align}
\label{eq:LiDAREqnCE}
\eta_{\scriptscriptstyle \text{LE}} = \rho \cdot \zeta_{\scriptscriptstyle \text{TX}}\cdot \zeta_{\scriptscriptstyle \text{RX}} \cdot  \frac{A(z)}{\pi z^2}   \cdot O(z) 
\end{align}
Here $\rho$ is the target reflectivity, $\zeta_{\scriptscriptstyle \text{RX}}\cdot \zeta_{\scriptscriptstyle \text{TX}}$ is the two-way system transmission efficiency,  $A(z)$ is the effective area of the receiver aperture, and $O(z)$ is the crossover or overlap function~\cite{Sassen82}.  
In both \eqref{eq:LiDAREqnCE} and \eqref{eqn:FSL_final}, $\rho$ is explicitly present and $\zeta$ exists due to normalization of $I_{\scriptscriptstyle \text{RX}},I_{\scriptscriptstyle \text{RX}}$ at their respective ports.
The distinction between models is therefore contained in the free-space loss terms, $\frac{A(z)}{\pi z^2} \cdot O(z) $, so we will outline their origin.  
\begin{figure}[ht]
\centering
\fbox{\includegraphics[width=0.95\linewidth]{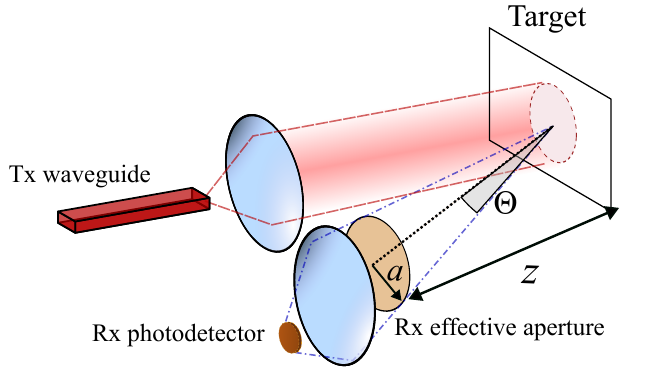}}
\caption{An illustration of a biaxial \mbox{LiDAR} configuration as modeled by the LiDAR Equation, where transmit (TX) waveguide and receive (RX) photodetector are collimated by separate optical systems. 
The reflected power is treated as orginating from a point source at the target, and the collected fraction of this power is that subtended by a projection of the photodetector at $z = 0$, an effective aperture of radius $a$.}
\label{fig:LiDAR_eqn}
\end{figure}

The area of the projected physical receiver at range $z = 0$, as perceived by a target at distance $z$, is the effective aperture area $A(z)$.
As depicted in Fig.~\ref{fig:LiDAR_eqn}, the reflecting target is then treated as the corresponding point source, with a Lambertian cosine radiation pattern aimed at a circular effective aperture of radius $a$.
The received power fraction can then be geometrically calculated from the half-angle, $\Theta = \atan(\frac{a}{z})$, that subtends the effective aperture.  
\begin{align}
\label{eq:subtended}
\frac{1}{\pi}{\iint_{\theta=0}^\Theta \cos \theta d\Omega }=\sin^2\Theta \approx \left( \frac{a}{z} \right )^2 \equiv \frac{A}{\pi z^2}
\end{align}
Here $d\Omega=\sin\theta d\theta d\phi$ is the solid angle infinitesimal element for polar and azimuthal spherical coordinates $\theta,\phi$. 
The radius $a(z)$ of the effective area $A(z)$ was calculated numerically for the {\it LiDAR Equation} curve in Fig.~\ref{fig:model_validation}(b) using a Zemax OpticStudio model of the Thorlabs F810APC-1550 doublet collimator.  
The method we used was as follows: a point source was placed on the optical axis at a distance $z$ in front of the lens and a \SI{9}{\micro\meter} diameter circulator aperture was placed on the optical axis behind the lens at a back focal length of \SI{3.08}{mm}.
The circular aperture here is being used to emulate the optical fiber core.   
We then traced the ray with the maximal angle from the point source that passed through the \SI{9}{\micro\meter} diameter circular aperture, and from this ray path we can record the maximum transverse offset of the ray from the optical axis at the front of the lens $z=0$.
 This transverse distance is used as the radius $a(z)$ of the effective area $A(z)$. 
Sweeping the distance $z$ then provides the values for $A(z)$ shown in Table~\ref{table:beep}. 

\onecolumngrid

\begin{table}[h!]
\centering
\vspace{0.5\baselineskip}
\begin{tabular}{|c|*{13}{c|}}

\hline
\textbf{$z$}, \SI{}{\meter}  & 3 & 4 & 5 & 6 & 7 & 8 & 9 & 10 & 11 & 12 & 13 & 14 & 15 \\
\hline
\textbf{$A(z)$}, \SI{}{\milli\meter^2} & 0.7 & 1.5 & 2.8 & 5.2 & 9.1 & 15.6 & 25.4 & 37.6 & 49.5 & 59.7 & 68.1 & 75.0 & 80.7 \\
\hline
\end{tabular}
\vspace{0.2\baselineskip}
\caption{The calculated values of the effective area $A(z)$ at the corresponding distances $z$, using the Zemax OpticStudio model of a Thorlabs F810APC-1550 doublet collimator with a back focal length of  \SI{3.08}{mm}.   This $A(z)$ data, in combination with $\rho=0.99$, $\zeta_{\scriptscriptstyle \text{TX}}=\zeta_{\scriptscriptstyle \text{RX}} =1$ and $O(z)=1$, is used with \eqref{eq:LiDAREqnCE} to generate the {\it LiDAR Equation} curve in Fig.~\ref{fig:model_validation}(b).}
\label{table:beep}
\end{table}

\twocolumngrid

The remaining term in the LiDAR Equation, the overlap function $O(z)$, is defined as the integrated fraction of the transmit power that intersects with the projection of the receiver’s aperture onto the target. 
This function represents a correction to \eqref{eq:subtended}, removing the fraction of transmitted power that the receiver is not capable of capturing.
It is worth mentioning that $O(z)$ behaves differently to the intensity overlap in~\eqref{eqn:FSL_final} given the normalization $\int I \;dA=1$; while \eqref{eqn:FSL_final} involves an integral of the product of intensities, $O(z)$ involves an integral of only a single intensity.  
As an example, when reducing the size of the TX spot, $O(z)$ is unaffected once the spot lies within the projected receive aperture and $O(z) = 1$, whereas \eqref{eqn:FSL_final} will continue to be impacted.  
A key prediction of~\eqref{eq:subtended} is that once the limiting physical aperture of the system is reached, typically at large $z$, the $1/z^2$ term will set the \mbox{LiDAR} system's optical power loss with increasing range. 

With the preceding information, we can identify three scenarios where we expect the predictions of the \mbox{LiDAR} Equation to differ from those of the model presented in this Article: 
\begin{itemize}[itemsep=0.5em]
\item[(i)]  Losses due to the mismatch of field distributions at the guided-mode receivers.
\item[(ii)]  Short distances where it is inaccurate to treat the reflection as radiation from a Lambertian point source at the target.
\item[(iii)]  Effects of wave propagation exist that are not accurately handled by geometric optics, such as diffraction.
\end{itemize}

\section{Simulating Expected Free-Space Loss}
\label{apnd:SimFSL}

\begin{figure}[h]
\centering
\fbox{\includegraphics[width=0.98\linewidth]{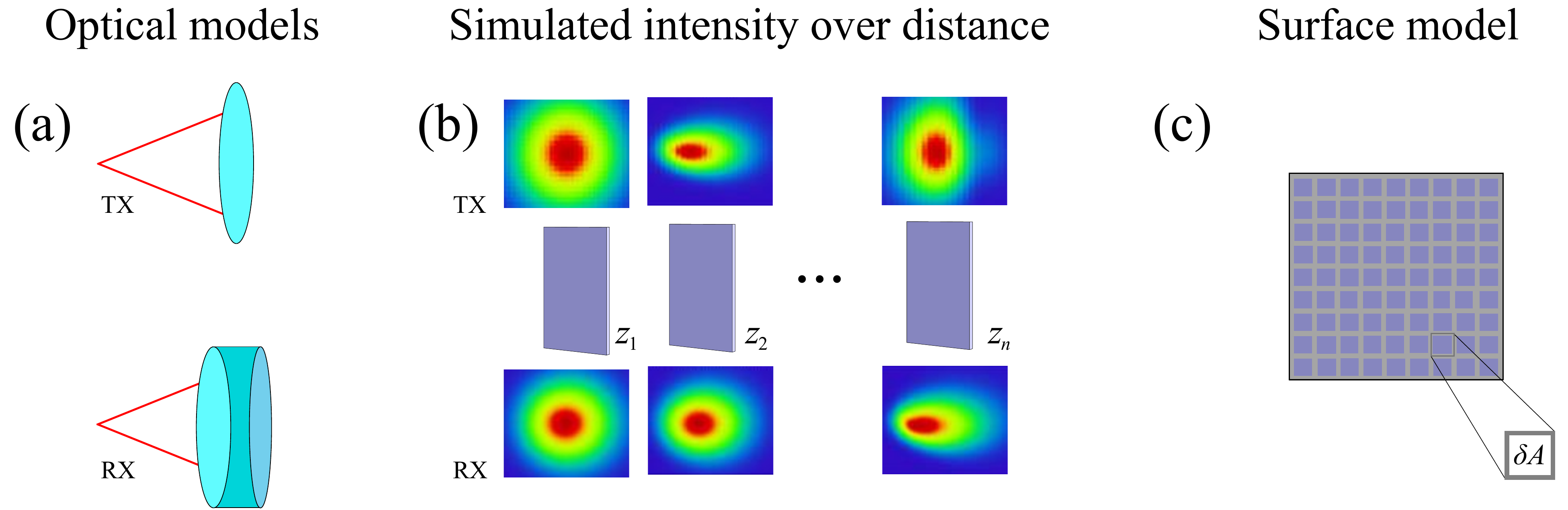}}
\caption{
Schematic of the key components used to generate the \textit{Model - Sim. Intensity} curves in both Figs.~\ref{fig:model_validation} and \ref{fig:predictive} using \eqref{eqn:FSL_final}.
(a) Optical models for the given transceiver are used to generate (b) simulated intensity profiles covering the desired target distances $z = z_1, z_2, \ldots, z_n$. (c) The surface model for the 99\% Spectralon\textsuperscript\textregistered target used in this work is then modeled by a known $\rho = 0.99$ and a calibrated $\delta A = \SI{1.20}{\micro\meter^2}$.
}
\label{fig:sim_methods}
\end{figure}

To clarify the procedure used to generate the \textit{Model - Sim. Intensity} curves in both Figs.~\ref{fig:model_validation} and \ref{fig:predictive}, we outline the steps as depicted in Fig.~\ref{fig:sim_methods}(a)-(c). The procedure consists of three main steps.
First, we build accurate simulation models of the given physical optics system (Fig.~\ref{fig:sim_methods}(a)). Next, we simulate the intensity of both transmitted beams and back-propagated receiver beams (Fig.~\ref{fig:sim_methods}(b)).
These intensity profiles, $I_{\scriptscriptstyle \text{TX}}$ and $I_{\scriptscriptstyle \text{RX}}$, are normalized such that there is unitary integrated intensity at the source, $\int I = 1$. 
Finally, we determine both the $\rho$ and the $\delta A$ that accurately represent the diffuse scattering surface (Fig.~\ref{fig:sim_methods}(c)).
In the case of the 99\% Spectralon\textsuperscript\textregistered target surface used in this work, we know $\rho=0.99$ and determine $\delta A = \SI{1.20}{\micro\meter^2}$ experimentally in Section~\ref{sec:model_calibration}.  
The associated $I_{\scriptscriptstyle \text{TX}}$, $I_{\scriptscriptstyle \text{RX}}$, $\rho$ and $\delta A$  terms are then inserted into \eqref{eqn:FSL_final} to obtain predictions for expected free-space loss that are shown in Figs.~\ref{fig:model_validation} and \ref{fig:predictive}.

\bibliography{Long_Range_Model_LIDAR}


\end{document}